\documentclass[twocolumn,showpacs,amsmath,amssymb]{revtex4}
\usepackage{graphicx}
\usepackage{dcolumn}
\usepackage{bm}
\usepackage{longtable}
\newcommand{\be}{\begin{eqnarray}}
\newcommand{\en}{\end{eqnarray}}
\newcommand{\ben}{\begin{eqnarray*}}
\newcommand{\enn}{\end{eqnarray*}}
\newcommand{\pa}{\partial}

\newcommand{\f}{\frac}

\newcommand{\bi}{\begin{itemize}}
\newcommand{\ei}{\end{itemize}}

\newcommand{\la}{\langle}
\newcommand{\ra}{\rangle}

\renewcommand{\r}{\rho}

\renewcommand{\d}{\delta}

\begin{document}
\title{On scaling laws in turbulent magnetohydrodynamic Rayleigh-Benard convection}
\author{Sagar Chakraborty}
\email{sagar@bose.res.in}
\affiliation{S.N. Bose National Centre for Basic Sciences\\Saltlake, Kolkata 700098, India}
\date{\today}
\begin{abstract}
We invoke the concepts of magnetic boundary layer and magnetic Rayleigh number and use the magnetic energy dissipation rates in the bulk and the boundary layers to derive some scaling laws expressing how Nusselt number depends on magnetic Rayleigh number, Prandtl number and magnetic Prandtl number for the simple case of turbulent magnetohydrodynamic Rayleigh-Benard convection in the presence of uniform vertical magnetic field.
\end{abstract}
\pacs{47.55.P-, 47.65.-d}
\maketitle
Turbulent fluid convection is an unsolved problem having very wide applications in the study of convective processes in atmospheres, oceans, metallurgy {\it etc.}
If the fluid is conducting and is acted upon by magnetic field then a theory for this magnetohydrodynamic fluid's convection hopes to explain, in the long run, the convection processes in planetary core, stellar interior and; some other important astrophysical, industrial and geophysical situations.
Complex nature of these realistic situations have prompted the researchers first to try to find a theory for the rather simpler problem of Rayleigh-Benard (RB) convection\cite{Chandrasekhar}, probably also because doing experiment on it is quite feasible.
\\
Briefly speaking, RB convection in MHD fluid is investigated in a set up where a magnetofluid of density $\rho$, kinematic viscosity $\nu$, conductivity $\sigma$, thermal diffusivity $\kappa$, magnetic permeability $\mu$ and isobaric thermal expansion coefficient $\alpha$ is confined between two horizontal plates (conducting or non-conducting).
Theoretically, the plates are considered to have infinite extent; in reality, this can be realised by making the thickness $d$ of magnetofluid between them very small compared to the lateral extent.
The plates are maintained at a constant relative temperature difference $\Delta$.
The constant acceleration due to gravity $\vec{g}$ is acting in the downward direction and the entire set up is acted upon by constant uniform vertical magnetic field $\vec{B}$.
Experiments\cite{Cioni, Aurnou} on MHD RB convection has been done using gallium or mercury confined between two copper plates.
However, whether the results of ref.\cite{Aurnou} applies to this paper is a question because this paper is applicable for high Nusselt number flows unlike the reference.
\\
If $\vec{u}(x,y,z,t)$, ${p}(x,y,x,t)$, ${T}(x,y,z,t)$ and $\vec{b}(x,y,z,t)$ be the velocity field, the kinematic pressure field (containing also the external divergence-free force terms), the temperature field and the magnetic field perturbation respectively, then the equations describing the convection under the Boussinesq's approximation are:
\be
&&\pa_tu_i+u_j\pa_ju_i=-\pa_ip+\nu\pa_j\pa_ju_i+\alpha gT\delta_{i3}\nonumber\\
&&\phantom{\pa_tu_i+u_j\pa_ju_i=}+\f{1}{\mu\rho}B_j\pa_jb_i\label{RB1}\\
&&\pa_tb_i+u_j\pa_jb_i=B_j\pa_ju_i+b_j\pa_ju_i+\f{1}{\sigma\mu}\pa_j\pa_jb_i\label{RB2}\\
&&\pa_tT+u_j\pa_jT=\kappa\pa_{j}\pa_jT\label{RB3}
\en
The second order terms have been purposefully retained in these three equations.
The boundary conditions at the two horizontal rigid plates, assumed perfectly conducting (electrically), are: a) $\vec{u}=0$ at $z=0$ and $z=d$, b) $T=\pm\Delta/2$ on $z=0$ and $d$ respectively and, c) $h_z=0$ on the plates.
As the convection becomes turbulent, assuming the existence of wind of turbulence {\it i.e.}, there exists a mean large scale velocity $U$ that stirs the magnetofluid in the bulk, we can define a Reynolds number $Re$ along with other important non-dimensional parameters.
These are listed below:
\ben
&&\textrm{Reynolds number: } Re=Ud/\nu\\
&&\textrm{Rayleigh number: } Ra=g\alpha\Delta d^3/\kappa\nu\\
&&\textrm{Chandrasekhar number: } Q=\sigma B^2d^2/\rho\nu\\
&&\textrm{Prandtl number: } Pr=\nu/\kappa\\
&&\textrm{Magnetic Prandtl number: } Pm=\mu\sigma\nu\\
&&\textrm{Nusselt number: } Nu=(\overline{u_zT}-\kappa\pa_3\overline{T})/(\kappa\Delta/d)
\enn
where the overline denotes average over any horizontal plane. By the way, in this paper we shall use angular brackets to denote volume average.
\\
It has been conjectured\cite{JKB} that there are various regimes in MHD RB convection wherein the scaling laws connecting $Nu$ with $Ra$ and $Q$ are different; this fact has been verified later by experiments\cite{Cioni}.
We shall take the results of these experiments at their face values although some deeper meaning behind them must exist; most importantly we shall make the following two observations from the results of that paper:
\begin{enumerate}
\item[(i)] The $Nu$ mostly depends on some power of the ratio $Ra/Q$. This ratio basically is a sort of Rayleigh number --- which we shall call magnetic Rayleigh number $Rb$ --- constructed with magnetic viscosity $\sigma B^2 d^2/\rho$ that essentially is the manifestation of Joule damping present in the magnetofluid.
\item[(ii)] It is surprising that $Nu-1$, rather than $Nu$, seems to be proportional to some powers of $Rb$.
\end{enumerate}
Of late, a paper\cite{Grossman} on turbulent RB convection in ordinary fluids has presented a unifying theory on how $Nu$ depends on $Ra$ by constructing four regimes each of which is defined depending on what (the boundary layers or the bulk of the fluid) the dominant contributors to the kinetic and the thermal dissipation rates are.
We shall closely follow this very paper extending the arguments and the assumptions to the turbulent RB convection in magnetofluid in the presence of uniform vertical magnetic field keeping in mind the observations (i) and (ii) made above.
\\
To begin with, we define $\varepsilon_u=\nu\la(\pa_iu_j)(\pa_iu_j)\ra$, $\varepsilon_T=\kappa\la(\pa_iT)(\pa_iT)\ra$ and $\varepsilon_b=(\mu^2\sigma\rho)^{-1}\la(\pa_ib_j)(\pa_ib_j)\ra$.
Using equations (\ref{RB1}) and (\ref{RB2}), and the definition of the $Nu$, we arrive at:
\be
\varepsilon_u+\varepsilon_b=\f{\nu^3}{d^4}\f{Ra}{Pr^2}(Nu-1)
\label{eueb}
\en
Similarly, usage of equation (\ref{RB3}) yields:
\be
\varepsilon_T=\kappa\left(\f{\Delta}{d}\right)^2Nu
\label{et}
\en
We shall now clearly state the main assumptions involved.
The very first and most important assumption of this paper is:
\be
\varepsilon_u\sim\f{\nu^3}{d^4}\f{Ra}{Pr^2}(Nu-1)
\label{a1}\\
\varepsilon_b\sim\f{\nu^3}{d^4}\f{Ra}{Pr^2}(Nu-1)
\label{a2}
\en
where (and henceforth, `$\sim$' means `scales as').
This explains (or is explained by) the observation (ii) above.
Secondly, we assume that there is a magnetic boundary layer (BL) of thickness $\d_b$ where the perturbation magnetic field is strongly affected by the magnetic diffusivity so that it grows to some definite value at the top of the layer from zero on the surface of the plate.
Thirdly, to go with the observation (i) above we shall churn this theory in such a manner that $Nu$ depends on $Rb$ and not on $Ra$ explicitly.
Fourthly, we shall confine ourselves to the laminar boundary layers only though the bulk of the magnetofluid is in turbulent state. So the thickness of the Blasius type kinetic BL $\d_u$ and the thermal boundary layer $\d_T$ are given by $\d_u\sim d/\sqrt{Re}$ and $\d_T\sim dNu^{-1}$ respectively.
Lastly, we shall confine our attention to the relevant cases where convection dominates conduction {\it i.e.}, $Nu\gg1$.
\\
Armed with these hypotheses, we propose that there may be following eight possible different regimes depending on whether the bulk or the respective boundary layers is chief contributor to dissipations:
\begin{enumerate}
\item[(A):] $\varepsilon_b$($bl$), $\varepsilon_T$($bl$), $\varepsilon_u$($bl$).
\item[(B):] $\varepsilon_b$($bl$), $\varepsilon_T$($bl$), $\varepsilon_u$($bulk$).
\item[(C):] $\varepsilon_b$($bl$), $\varepsilon_T$($bulk$), $\varepsilon_u$($bl$).
\item[(D):] $\varepsilon_b$($bl$), $\varepsilon_T$($bulk$), $\varepsilon_u$($bulk$).
\item[(E):] $\varepsilon_b$($bulk$), $\varepsilon_T$($bl$), $\varepsilon_u$($bl$).
\item[(F):] $\varepsilon_b$($bulk$), $\varepsilon_T$($bl$), $\varepsilon_u$($bulk$).
\item[(G):] $\varepsilon_b$($bulk$), $\varepsilon_T$($bulk$), $\varepsilon_u$($bl$).
\item[(H):] $\varepsilon_b$($bulk$), $\varepsilon_T$($bulk$), $\varepsilon_u$($bulk$).
\end{enumerate}
Here, `($bl$)' and `($bulk$)' respectively says that the respective boundary layer and the bulk is dominant contributor to the dissipation concerned.
These regimes can be prepared by proper choice of $Ra$, $Pr$ and $Pm$.
For example, for large $Ra$ regime (H) is possible; for small $Pr$ and large $Pm$ regime (F) is expected and so on.
\\
Before proceeding further, let us estimate the width of the magnetic boundary layer.
For $Pm\ll1$ ({\it i.e.}, $\d_u\ll\d_b$), we may neglect the first and the second terms in the R.H.S. of the equation (\ref{RB2}) and under steady convection one is left to compare the second term of L.H.S. to the only unneglected term in the R.H.S:
:
\be
&&\left|u_j\pa_jb_i\right|\sim\left|\f{1}{\mu\sigma}\pa_j\pa_jb_i\right|\\
\Rightarrow&&U\f{B}{\d_b}\sim\f{B}{\mu\sigma\d_b^2}\\
\Rightarrow&&\d_b\sim\f{d}{RePm}
\label{mblw1}
\en
Note that while estimating gradients we are using the magnitude of the uniform external magnetic field $B$ rather than the typical magnetic field fluctuations. %
This also is an assumption in analogy with the standard assumption made in ordinary fluid RB convection where while estimating the temperature gradients $\Delta$ is used freely.
For $Pm\gg1$ ({\it i.e.}, $\d_u\gg\d_b$), we must balance the first and the third term in the R.H.S. of the equation (\ref{RB2}):
\be
&&\left|B_j\pa_ju_i\right|\sim\left|\f{1}{\mu\sigma}\pa_j\pa_jb_i\right|\\
\Rightarrow&&B\f{U}{\d_u}\sim\f{B}{\mu\sigma\d_b^2}\\
\Rightarrow&&\d_b\sim\f{d}{Re^{\f{3}{4}}Pm^{\f{1}{2}}}
\label{mblw2}
\en
where the gradient of the velocity field in the kinematic boundary layer has been approximated by a linearly increasing profile.
\\
Now under the assumptions made, following can be easily derived
\be
\varepsilon_u(bulk)\sim\f{U^3}{d}=\f{\nu^3}{d^4}Re^3\label{eub}\\
\textrm{For }Pr\ll1,\phantom{x}\varepsilon_T(bulk)\sim\f{U\Delta^2}{d}=\kappa\left(\f{\Delta}{d}\right)^2PrRe\label{etb1}\\
\textrm{For }Pr\gg1,\phantom{x}\varepsilon_T(bulk)\sim\f{U\Delta^2}{d}\f{\d_T}{\d_u}\nonumber\\
\phantom{\textrm{For }Pr\gg1,\phantom{x}\varepsilon_T(bulk)}\sim\kappa\left(\f{\Delta}{d}\right)^2\f{PrRe^{\f{3}{2}}}{Nu}\label{etb2}\\
\textrm{For }Pm\ll1,\phantom{x}\varepsilon_b(bulk)\sim\f{B^2}{2\mu\r}\f{U}{d}\sim\f{\nu^3}{d^4}\f{QRe}{Pm}\label{ebb1}\\
\textrm{For }Pm\gg1,\phantom{x}\varepsilon_b(bulk)\sim\f{B^2}{2\mu\r}\f{U}{d}\f{\d_T}{\d_u}\sim\f{\nu^3}{d^4}\f{QRe^{\f{3}{4}}}{Pm^{\f{3}{2}}}\label{ebb2}\\
\varepsilon_u(bl)\sim\nu\left(\f{U}{\d_u}\right)^2\f{\d_u}{d}\sim\f{\nu^3}{d^4}Re^{\f{5}{2}}\label{eubl}
\en
Though we are interested in the limit of large and small Prandtl numbers (magnetic and non-magnetic), they can not be too large or too small, for, then either the convective flow will be suppressed or the thermal and the magnetic diffusivities will be so high that the respective effects will be the dominance of thermal conduction and too quick decay of magnetic fields to have any magnetic effect on the fluid.
In writing the equation (\ref{ebb1}), it has been kept in mind that the magnetic energy per unit mass $B^2/2\mu\r$ is cascaded down the spatial scales by the turbulent eddies until it is dissipated at the lowest scale (the turbulent condition is taken to be stationary with continuous feeding of magnetic energy into the magnetofluid) and in the equation (\ref{eubl}), the factor $\d_u/{d}$ comes in to assert that only the kinematic boundary layers' volume is involved.
Attempt to get similar relations for $\varepsilon_T(bl)$ takes one to the expression (\ref{et}).
Therefore, alternate relations must be sought.
Using the equation (\ref{RB3}) and the arguments given in the Landau's text on fluid mechanics\cite{Landau}, one readily arrives at:
\be
\textrm{For }Pr\ll1,\phantom{x}Nu\sim Re^{\f{1}{2}}Pr^{\f{1}{2}}\label{etbl1}\\
\textrm{For }Pr\gg1,\phantom{x}Nu\sim Re^{\f{1}{2}}Pr^{\f{1}{3}}\label{etbl2}
\en
Again, coming on the scaling expression for $\varepsilon_b(bl)$ we work out as follows:
In the magnetic boundary layer if $\vec{j}$ is the current density, then from the relation $\vec{\nabla}\times\vec{b}=\mu \vec{j}$ one can estimate $j$ as $B/\d_b\mu$.
This means that mean rate of dissipation of magnetic energy per unit mass $j^2/\sigma\r$ is given as:
\be
\varepsilon_b(bl)=\f{B^2}{\sigma\r\mu^2\d_b^2}
\label{ebbl12}
\en
This equation alongwith with relations (\ref{mblw1}) and (\ref{mblw2}) gives $\varepsilon_b(bl)$ for $Pm\ll1$ and $Pm\gg1$ respectively.
\\
Now we are prepared to derive a range of scaling laws.
\begin{table}[ht]
\caption{For $Pm\ll1$ and $Pr\ll1$. (UD=Undetermined)}
\centering
\begin{tabular}{||c||c|c|c||}
\hline\hline
Regime&Exponent of $Rb$&Exponent of $Pr$ & Exponent of $Pm$\\\hline\hline
A&$\f{1}{3}$&0&0\\\hline
B&$\f{1}{3}$&0&0\\\hline
C&UD&UD&UD\\\hline
D&UD&UD&UD\\\hline
E&1&-1&1\\\hline
F&1&-1&1\\\hline
G&UD&UD&UD\\\hline
H&UD&UD&UD\\\hline\hline
\end{tabular}
\label{table1}
\end{table}
%
\begin{table}[ht]
\caption{For $Pm\ll1$ and $Pr\gg1$. (UD=Undetermined)}
\centering
\begin{tabular}{||c||c|c|c||}
\hline\hline
Regime&Exponent of $Rb$&Exponent of $Pr$ & Exponent of $Pm$\\\hline\hline
A&$\f{1}{3}$&$-\f{2}{9}$&0\\\hline
B&$\f{1}{3}$&$-\f{2}{9}$&0\\\hline
C&$\f{3}{5}$&$-\f{2}{5}$&0\\\hline
D&$\f{3}{5}$&$-\f{2}{5}$&0\\\hline
E&1&$-\f{4}{3}$&1\\\hline
F&1&$-\f{4}{3}$&1\\\hline
G&3&-4&3\\\hline
H&3&-4&3\\\hline\hline
\end{tabular}
\label{table2}
\end{table}
%
\begin{table}[ht]
\caption{For $Pm\gg1$ and $Pr\ll1$. (UD=Undetermined)}
\centering
\begin{tabular}{||c||c|c|c||}
\hline\hline
Regime&Exponent of $Rb$&Exponent of $Pr$ & Exponent of $Pm$\\\hline\hline
A&$\f{1}{2}$&$-\f{1}{8}$&$\f{1}{2}$\\\hline
B&$\f{1}{2}$&$-\f{1}{8}$&$\f{1}{2}$\\\hline
C&UD&UD&UD\\\hline
D&UD&UD&UD\\\hline
E&2&$-\f{5}{2}$&3\\\hline
F&2&$-\f{5}{2}$&3\\\hline
G&UD&UD&UD\\\hline
H&UD&UD&UD\\\hline\hline
\end{tabular}
\label{table3}
\end{table}
%
\begin{table}[ht]
\caption{For $Pm\gg1$ and $Pr\gg1$. (UD=Undetermined)}
\centering
\begin{tabular}{||c||c|c|c||}
\hline\hline
Regime&Exponent of $Rb$&Exponent of $Pr$ & Exponent of $Pm$\\\hline\hline
A&$\f{1}{2}$&$-\f{1}{2}$&$\f{1}{2}$\\\hline
B&$\f{1}{2}$&$-\f{1}{2}$&$\f{1}{2}$\\\hline
C&1&-1&1\\\hline
D&1&-1&1\\\hline
E&2&-3&3\\\hline
F&2&-3&3\\\hline
G&UD&UD&UD\\\hline
H&UD&UD&UD\\\hline\hline
\end{tabular}
\label{table4}
\end{table}
%
We shall demonstrate the line of attack for doing so by finding out a illustrative scaling law for the regime (A); all other seven regimes can be treated similarly.
For the regime (A) the dominant contributor to the magnetic energy dissipation are the magnetic BLs.
Deciding to work for $Pm\gg1$ first and as $Nu\gg1$, we equate the corresponding relation (\ref{ebbl12}) (using the expression (\ref{mblw2}) for $\d_b$) to the expression (\ref{a2}) to get:
\be
&&\f{\nu^3}{d^4}RaNuPr^{-2}\sim \f{B^2}{\sigma\r\mu^2\d_b^2}\\
\Rightarrow&&Nu\sim\f{1}{Rb}\left(\f{Pr}{Pm}\right)^2\left(\f{d}{\d_b}\right)^2\\
\Rightarrow&&Nu\sim\f{1}{Rb}\f{Pr^2}{Pm}Re^{\f{3}{2}}
\label{qq1}
\en
Again in the regime (A), the dominant contributors to the kinetic energy dissipation are the kinetic BLs so we equate relation (\ref{eubl}) to relation (\ref{a1}) to get:
\be
Re^{\f{5}{2}}=RaNuPr^{-2}
\label{qq2}
\en
Also, they are the BLs --- thermal BLs --- that are contributing to the thermal energy dissipation, so relation (\ref{etbl1}) is of importance (we are considering $Pr\ll1$ to begin with) for this case.
For other cases, where the contribution to the thermal energy dissipation comes from the bulk, the corresponding $\varepsilon_T(bulk)$ will have to be compared with equation (\ref{et}) to obtain the desired relations.
Substituting relation (\ref{etbl1}) into relation (\ref{qq2}), we obtain:
\be
Re\sim Ra^{\f{1}{2}}Pr^{-\f{3}{4}}
\label{qq3}
\en
which in turn when put into relation (\ref{qq2}), yields:
\be
Nu\sim Ra^{\f{1}{4}}Pr^{-\f{1}{8}}
\label{qq4}
\en
Now, in accordance with the observation (i), we eliminate $Ra$ from relations (\ref{qq3}) and (\ref{qq4}) to write $Re$ in terms of $Nu$ and $Pr$.
This is going to be the general strategy throughout.
Thus, for this example we have:
\be
Re\sim Nu^{2}Pr^{-1}
\label{qq5}
\en
Putting relation (\ref{qq5}) in relation (\ref{qq1}), we arrive at the desired scaling law:
\be
Nu\sim Rb^{\f{1}{2}}Pr^{-\f{1}{8}}Pm^{\f{1}{2}}
\label{qq6}
\en
The benefit of this strategy of finding scaling laws is that the exponents of $Pr$ and $Pm$ are also predicted.
Moreover, the method of classification of the regimes in the present-day experiments\cite{Cioni} differs from what has been done theoretically here.
Hence, each of those regimes may consist of an overlapping region of some of the eight regimes proposed in this paper.
This would suggest that a fit of the form $Nu=cRa^a/Q^b$ ($a,b,c$ are just numerical constants) usually done to represent the experimental results can in principle be an equally good fit like: $Nu=c_1Rb^{d_1}+c_2Rb^{d_2}+c_3Rb^{d_3}+\cdots$; this of course is not mathematically impossible.
\\
In closing, we list rest of the all mathematically possible scaling results (assuming $Nu\sim Rb^aPr^bPm^c$) in four tables --- table-\ref{table1} to table-\ref{table4} --- without showing explicit derivation whose strategy, anyway, has already been clearly outlined; and hope that experiments and simulations would be done in near future to test the conjecture proposed in this paper and the exponents of Prandtl number and magnetic Prandtl number will be concentrated upon more seriously.
Only then one can say if the idea presented herein, which seems to have a flavour of being a mere translation of the Grossmann-Lohse theory\cite{Grossman} of thermal flow in Rayleigh-Benard geometry for similar convection processes in magnetohydrodynamic flows, is valid or not.
We remind the readers that it has not been possible for us to investigate the scaling laws when the laminar BLs turn turbulent: This, of course, should draw attention of the theoretical fluid dynamists.
\\
\\
The author would like to acknowledge his supervisor --- Prof. J. K. Bhattacharjee --- for the helpful and fruitful discussions.
Also, CSIR (India) is gratefully acknowledged for awarding fellowship to the author.

%
\end{document}